\documentstyle[aas2pp4,flushrt]{article}	

\received{}
\accepted{}
\slugcomment{Submitted to {\it The Astrophysical Journal}}
\lefthead{Lauer et al.}
\righthead{The Far Field Hubble Constant}

\def\viz{\ifmmode(V{-}I)_0\else$(V{-}I)_0$\fi}
\def\mbar{\ifmmode\overline{m}\else$\overline{m}$\fi}
\def\Mbar{\ifmmode\overline{M}\else$\overline{M}$\fi}
\def\Mibar{\ifmmode\overline{M}_{I_{\rm KC}}\else$\overline{M}_{I_{\rm KC}}$\fi}\def\f8bar{\ifmmode\overline{I}_{\rm 814}\else$\overline{I}_{\rm 814}$\fi}
\def\Mf8bar{\ifmmode\overline{M}_{I_{\rm 814}}\else$\overline{M}_{I_{\rm 814}}$\fi}
\def\s.{\kern+ .1em\hbox{$\buildrel ^{\prime\prime} \over 
    {\rm .}$}} 
 
\begin{document}
\title{The Far Field Hubble Constant\altaffilmark{1}} 

\author{Tod R. Lauer}
\affil{Kitt Peak National Observatory, National Optical Astronomy
Observatories,\altaffilmark{2} P.~O. Box 26732, Tucson, AZ 85726}
\affil{Electronic mail: lauer@noao.edu}

\author{John L. Tonry}
\affil{Institute for Astronomy, University of Hawaii, 2680 Woodlawn
Dr., Honolulu, HI 96822}
\affil{Electronic mail: jt@avidya.ifa.hawaii.edu}

\author{Marc Postman\altaffilmark{3}}
\affil{Space Telescope Science Institute,\altaffilmark{4} 3700 San
Martin Dr., Baltimore, MD 21218}
\affil{Electronic mail: postman@stsci.edu}

\author{Edward A. Ajhar}
\affil{Kitt Peak National Observatory, National Optical Astronomy
Observatories,\altaffilmark{2} P. O. Box 26732, Tucson, AZ 85726}
\affil{Electronic mail: ajhar@noao.edu}

\and

\author{Jon A. Holtzman}
\affil{New Mexico State University, Box 30001, Dept. 4500, Las Cruces,
NM 88003}
\affil{Electronic mail: holtz@nmsu.edu}

\altaffiltext{1}{Based on observations with the NASA/ESA {\it Hubble
Space Telescope,} obtained at the Space Telescope Science Institute
(STScI), which is operated by the Association of Universities for
Research in Astronomy (AURA), Inc., under National Aeronautics and
Space Administration (NASA) Contract NAS 5-26555.}
\altaffiltext{2}{The National Optical Astronomy Observatories are
operated by AURA, Inc., under cooperative agreement with the National
Science Foundation.}
\altaffiltext{3}{Visiting Astronomer Cerro Tololo Inter-American Observatory,
NOAO.}
\altaffiltext{4}{STScI is operated by AURA, Inc., under contract to
NASA.}

\begin{abstract}

We used {\it HST} to obtain surface brightness fluctuation (SBF) observations
of four nearby brightest cluster galaxies (BCG)
to calibrate the BCG Hubble diagram of \markcite{lp92} Lauer \& Postman (1992).
This BCG Hubble diagram contains 114 galaxies covering the full celestial sphere
and is volume limited to 15,000 km s$^{-1},$
providing excellent sampling of the far-field Hubble flow.
The SBF zero point is based on the Cepheid
calibration of the ground $I_{KC}$ method 
\markcite{t97}(Tonry et al. 1997) as extended to the WFPC2 F814W filter
by \markcite{a97} Ajhar et al. (1997).
The BCG globular cluster luminosity functions
give distances essentially identical to the SBF results.
Using the velocities and SBF distances of the four BCG alone
gives $H_0=82\pm8{\rm ~km~s^{-1}~Mpc^{-1}}$ in the CMB frame, valid
on $\sim$4,500 km s$^{-1}$ scales.
Use of BCG as photometric redshift estimators
allows the BCG Hubble diagram to be calibrated independently of
recession velocities, yielding a far-field
$H_0=89\pm10{\rm ~km~s^{-1}~Mpc^{-1}}$
with an effective depth of $\sim$11,000 km s$^{-1}$.
The error in this case is dominated by the photometric cosmic scatter in
using BCG as distance estimators.
The concordance of the present results with other recent $H_0$ determinations,
and a review of theoretical treatments on 
perturbations in the near-field Hubble flow, argue
that going to the far-field removes an important source
of uncertainty, but that there is not a large systematic error
to be corrected for to begin with.
Further improvements in $H_0$ depend more on understanding nearby
calibrators than on improved sampling of the distant flow.

\end{abstract}


\section{Introduction}

A key part of measuring the Hubble constant, $H_0,$
is to look out far enough so that the bulk velocities
of galaxies are trivial compared to the Hubble flow itself.
Due to the Virgo cluster infall pattern,
observation of the unbiased Hubble flow can only be contemplated
at distances in excess of $\sim3000$ km s$^{-1}$.
Furthermore, bulk flows on even larger scales,
such as those associated with the Great Attractor,
may bias measurement of $H_0.$
\markcite{tco92}Turner, Cen, \& Ostriker (1992) and
\markcite{shi}Shi, Widrow, \& Dursi (1996), for example, show that under
standard theories of structure formation, measurements
of $H_0$ can depart significantly from its true ``global'' value
due to the inhomogeneous distribution of matter in the universe,
unless care is taken to sample deeply with large angular coverage.
Indeed, a common concern with many recent $H_0$ determinations
is that they are not truly sampling the distant Hubble flow
\markcite{bar}(Bartlett et al. 1995).

Characterizing the far-field requires observing large numbers of
objects at large distances so that the
Hubble diagrams are insensitive to random peculiar velocities or bulk flows.
Hubble diagrams at present are largely based on
the Tully-Fisher or $D_n-\sigma$ relationships, the luminosities
of supernovae (SN Ia or SN II), and brightest cluster galaxies (BCG).
Tully-Fisher distances are available out to $\sim$9,000 km s$^{-1},$
and have been recently used to measure a far-field $H_0$
\markcite{gio97}(Giovanelli et al, 1997), while $D_n-\sigma$ have
full-sky coverage out to only $\sim$6,000 km s$^{-1}.$
Only a few SN II have been observed in sufficient detail at large distances
\markcite{schm}(Schmidt et al. 1994),
but the SNIa Hubble diagram is becoming richer with time
and provides some sampling of the Hubble flow
out to $\sim$30,000 km s$^{-1}$ \markcite{rpk}(Riess, Press, \&
Kirshner 1996; \markcite{ham} Hamuy et al. 1996).
At present, however, calibration of the SNIa distance scale
remains controversial (see \markcite{sand96}Sandage et al. 1996),
and the SN Ia diagrams remain relatively sparse at large distances.
In this work we focus on calibrating the BCG Hubble diagram,
which is based on a recent characterization of BCG as relative
distance estimators \markcite{pl}(Postman \& Lauer 1995).
 
\begin{figure}[tb]
\plotone{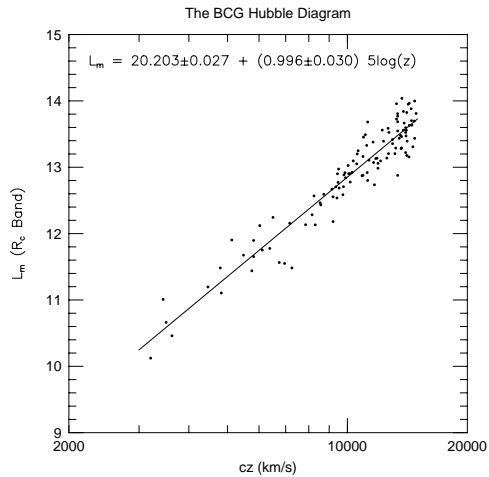}
\caption{The BCG Hubble diagram.  R-band metric
luminosities of the BCG, corrected by the $L_m-\alpha$ relationship,
are plotted as a function of velocity in the Local Group frame.
The line is the mean Hubble relation fitted.}
\label{bcg_hub}
\end{figure}

In the classic work of \markcite{sand72} Sandage (1972)
and \markcite{sand73} Sandage \& Hardy (1973),
BCG were used to show that the Hubble
flow was linear over a large range in redshift.
\markcite{lp} Lauer \& Postman (1994) observed BCG
to define a frame for measuring the peculiar velocity
of the Local Group, but as this work was in progress they realized that they
could test for $H_0$ variations with distance with greater precision
than was previously available in response to the concerns
of \markcite{tco92}Turner, Cen, \& Ostriker (1992).
\markcite{lp92} Lauer \& Postman (1992) presented a Hubble diagram
based on the 114 BCG that defined the
volume-limited full-sky sample of Abell clusters within 15,000 km s$^{-1},$
which is shown again here in Figure \ref{bcg_hub}.
In brief, the absolute magnitudes of BCG, $L_m,$ measured in apertures
of fixed metric size, $r_m,$ can be predicted from
$\alpha\equiv d\log L_m/d\log r|_{r_m}$ \markcite{hoe}(Hoessel 1980).
Figure \ref{bcg_hub} shows the metric luminosities as apparent fluxes,
corrected by the $L_m-\alpha$ relationship to a standard
value of $\alpha=0.5.$

The BCG Hubble diagram slope is $0.996\pm0.030$ of the
expected value, consistent with a uniform Hubble flow over $0.01\leq
z\leq0.05.$ \markcite{lp92} Lauer \& Postman (1992) limit
any variation in the {\it apparent} or local $H_0$
(the Hubble constant measured over a limited depth)
as compared to $H_0$ measured globally over the entire volume,
to $\delta_H\equiv\Delta H_0/H_0<0.07.$
The SNIa Hubble diagram also shows no evidence for $H_0$ variations
with distance; \markcite{rpk}Riess, Press, \& Kirshner (1996)
show its slope (relative to Euclidean) to be $1.005\pm0.018.$
The full-sky coverage of the Abell cluster sample is
crucial, as any dipole pattern caused by large bulk
flows (such at that advanced by \markcite{lp}Lauer \& Postman 1994)
will integrate out of the Hubble diagram to first order.
The linearity of the BCG Hubble diagram shows that an
excellent estimate of the far-field $H_0$ can be obtained once the
zero point of the diagram is calibrated.
We note that BCG presently provide the only volume-limited sample
that explores the Hubble flow at these distances.

A Hubble constant can be obtained from the BCG Hubble diagram
once an absolute distance is known to a subset of the galaxies.
In essence, one transfers the full sample to a common distance,
and finds the average absolute luminosity of the BCG on the
assumption that the calibrating set is typical.
Random velocities and bulk flows
of the BCG contribute to the ``cosmic scatter'' in their luminosity
distribution, but cause no systematic offset (with the caveats
discussed in $\S$\ref{far_enough}).
We contrast this approach to others that use the apparent
distance ratio between the Virgo and Coma clusters, or any other near
and far aggregate of galaxies, to reach the far-field.
Instead, we are using the BCG as complete probes of the Hubble flow
over a large volume.

We chose to calibrate the BCG Hubble diagram with surface
brightness fluctuation (SBF) distance estimates to four of the nearest BCG.
The SBF method \markcite{ts88} (Tonry \& Schneider 1988) uses the ratio of the
second to first moments of the
stellar luminosity function within early-type stellar systems
as a distance estimator.
The ratio of moments corresponds to an apparent
magnitude, \mbar, that in the near-IR corresponds to the
brightness of a typical red giant star.
When the images are deep enough such that a star of apparent luminosity
\mbar~contributes more than a single photon to an observation, the random
spatial point-to-point surface brightness fluctuations in a galaxy image
are dominated by the finite number of stars it comprises, rather
than photon shot noise.
A power spectrum of the SBF pattern provides \mbar.
Use of the SBF method on galaxies with distances known from other methods
\markcite{jetal92}(see Jacoby et al. 1992 for additional details)
provides the zero point \Mbar, allowing absolute distances
to be computed from \mbar.

The most recent calibration of the SBF method is presented by
\markcite{t97} Tonry et al. (1997).
Major components of this work are: 1) understanding
how \Mbar\ varies with stellar population,
2) determining the zero point of the method,
and 3) establishing the universality of the calibration.
Tonry et al. observe in the $I_{KC}$ band, which minimizes variations
in \Mibar with stellar population {\it ab initio.}
They also show that variations in \Mibar~are fully characterized
by the ($V-I$) colors of the stellar systems.
Based on 149 nearby galaxies they find
\begin{equation}
\label{eqMIbar}
\Mibar = (-1.74 \pm 0.07) + (4.5 \pm 0.25) [\viz - 1.15].
\end{equation}
This relationship has scatter of only 0.05 mag and agrees well
with the theoretical calculations of \markcite{wora} Worth\-ey (1993a,
\markcite{worb} 1993b) both in slope {\it and} zero point.
\markcite{ta92} Tammann (1992) was concerned that an earlier SBF calibration
based on $(V-I)$ was incomplete and that \Mibar~additionally depended
on the galaxies' ${\rm Mg_2}$ indices.
In response, Tonry et al. use their extensive sample to show that
there is no correlation between the residuals of equation (\ref{eqMIbar})
and ${\rm Mg_2.}$
The zero point of equation (\ref{eqMIbar}) is based on Cepheid distances
to seven spiral galaxies with bulge SBF observations.
Tonry et al. present numerous comparisons
of SBF to PNLF, Tully-Fisher, $D_N-\sigma$, SNIa, and SN II distances,
finding no evidence for any systematic offset between SBF bulge and elliptical
galaxy measurements, nor any other systematic effect that
challenges the calibration.

Although the nearest BCG are too far
away for the SBF method to work from the ground,
the high spatial resolution of {\it HST} allows
SBF to be used beyond the 15,000 km s$^{-1}$ depth of the
\markcite{lp} Lauer \& Postman (1994) sample.
An important caveat is that there is no direct
match to the $I_{KC}$ filter among the WFPC2 filter set.
The F814W filter is a close analogue to $I_{KC}$
(see \markcite{holtza}Holtzman et al. 1995a), but requires additional
calibration to tie it to the \markcite{t97}Tonry et al. (1997) zero point.
\markcite{a97} Ajhar et al. (1997) accomplished this task in preparation
for the present work, by comparing {\it HST} F814W SBF observations to the
$I_{KC}$ results for 16 galaxies in the Tonry et al. sample.
For the WFPC2 CCDs and F814W filter, Ajhar et al. find
\begin{equation}
\label{eqM8bar}
\Mf8bar = (-1.73 \pm 0.07) + (6.5 \pm 0.7) [\viz - 1.15],
\end{equation}
with scatter similar to that about equation (\ref{eqMIbar}).
A key difference between equation (\ref{eqMIbar}) and (\ref{eqM8bar})
is the steeper relationship between \Mf8bar and $(V-I),$
which Ajhar et al. show is consistent with the differences
between the F814W and $I_{KC}$ filters.
Calibration of {\it HST} for SBF work is thus crucial for the present work.

\section{Observations and Reductions}

\subsection{BCG Sample Selection}

\begin{figure}[tb]
\plotone{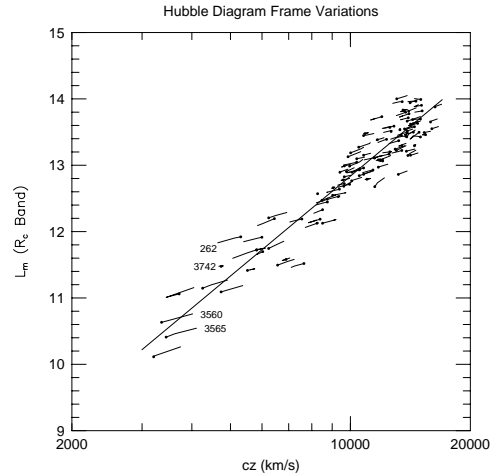}
\caption{The BCG Hubble diagram with frame variations.
This figure shows how the BCG move within the Hubble
diagram as the velocity frame changes.
The lines start at the positions of the BCG in the
CMB frame, move through the Local Group frame, ending
with the solid points in the Abell Cluster frame.
The four BCG with SBF distances are labeled.}
\label{bcg_hub_frame}
\end{figure}

We selected the BCG in four Abell clusters, A262, A3560, A3565, and
A3742 for observation.
These are among the nearest of the
\markcite{lp}Lauer \& Postman (1994) sample
so as to minimize {\it HST} exposure time.
We also wanted to minimize the effects of
bulk flows on placement of the calibrating BCG within the Hubble diagram,
so we selected BCG positioned such that their mean
photometric offset about the Hubble line in Figure \ref{bcg_hub} is largely
insensitive to whether the Hubble diagram is constructed from velocities
referenced to the cosmic microwave background
(CMB), Local Group (LG), or the Abell Cluster (AC) frame
solution of \markcite{lp}Lauer \& Postman (1994).
Figure \ref{bcg_hub_frame} shows how the positions of the BCG in the Hubble
diagram change with changing velocity system.
Complete frame invariance of the results is difficult
to achieve with only four BCG, however;
we discuss this issue later where our results are affected by it.

The BCG properties are given in Table \ref{tabsamp}.
Photometry, details of BCG identification, and so on are
discussed by \markcite{pl}Postman \& Lauer (1995).
Velocities are given in the CMB, LG, and AC frames
(see \markcite{lp}Lauer \& Postman 1994).
The velocities are a weighted average of all galaxies selected
to be within the given Abell cluster;
the estimated velocity error is 184 km s$^{-1}$
(see \markcite{pl}Postman \& Lauer 1995).
Extinctions are from \markcite{bh}Burstein \& Heiles (1984).

\begin{deluxetable}{rrrcccccr}
\tablecolumns{9}
\tablewidth{0pt}
\tablecaption{BCG Galaxy Sample and Observations}
\tablehead{
 &\multicolumn{2}{c}{J2000.0}   & \colhead{$V_{CMB}$} & \colhead{$V_{LG}$} &
\colhead{$V_{AC}$}& &  & \colhead{Time} \\
\colhead{Abell}& \colhead{RA} &
\colhead{DEC} & \colhead{${\rm km~s^{-1}}$} & \colhead{${\rm km~s^{-1}}$} &
\colhead{${\rm km~s^{-1}}$}& \colhead{${\rm E_{B-V}}$}    & \colhead{Date} &
\colhead{s} }
\startdata
 262& 01 52 46.3& +36 09 05 &4650& 5130 &5310& 0.060 & 96/02/11 & 16,400\nl
3560& 13 31 53.3& $-$33 14 04 &4020& 3510 &3360& 0.038 & 96/01/16 & 9,200\nl
3565& 13 36 39.1& $-$33 57 57 &4110& 3630 &3450& 0.030 & 96/01/19 & 11,600\nl
3742& 21 07 52.3& $-$47 10 43 &4680& 4800 &4740& 0.018 & 96/04/22 & 16,500\nl
\enddata
\tablecomments{Velocities are weighted averages of all available
cluster data given in the Cosmic Microwave Background (CMB), Local Group
at rest (LG), or Abell Cluster (AC) frames (see Lauer \& Postman 1994 and
Postman \& Lauer 1995).  ${\rm E_{B-V}}$ values are from Burstein \& Heiles
(1984).  The last two columns are date of the {\it HST} observations
and total exposure time.}
\label{tabsamp}
\end{deluxetable}

\subsection{{\it HST} Observations and Reductions}

Images were obtained in WFPC2 using the F814W filter.
The galaxies were centered in the high-resolu\-tion PC1 chip.
While suitable data were obtained in the flanking WFC CCDs, we chose
to analyze the PC data only, given its superior resolution,
and the greater brightness of the central portions of the BCG with
respect to the sky.
At the BCG distances, the apparent luminosity of an F814W SBF ``star,''
\f8bar, is extremely faint, thus long exposures are required;
the total exposures are given in Table \ref{tabsamp}.
Although our ideal criterion is to obtain at least five photons per
\f8bar star, our data contained only 2.3--3.8 photons per \f8bar.  The reasons
for the shortfall were 1) {\it HST} was 5\%\ less sensitive through F814W
than prelaunch numbers suggested, 2) the galaxies were about 5\%\ more
distant than we had guessed from their redshifts, and 3) \Mf8bar\ was
significantly fainter (0.65 mag in the case of A262) because the
galaxies were redder than anticipated and \f8bar was more sensitive to
color than our assumptions.  Nevertheless, all four galaxies yielded a
strong SBF signal that could be accurately determined.

Because compact artifacts can strongly affect the SBF power spectrum,
we built the total exposures from sets of ``dithered'' half-orbit
images to eliminate hot-pixels, CCD defects, as well as cosmic-ray hits.
Each individual exposure was typically 1200s long, with the actual
exposure time set to maximize the total exposure obtained with
two roughly equal exposures per orbit.
The dither pattern consisted of moving the telescope between exposures
in a skewed-square-spiral pattern, designed to achieve
integral-pixel shifts in the WFC CCDs.
The pattern optimized removal of fixed-pattern CCD defects,
while preserving the exact shape of any SBF pattern from exposure to exposure.
While one might argue for an approach that would instead optimize
information recovered with sub-pixel stepping, we did not have
the exposure time available to obtain an equal number of the required
$2\times2$ 0.5-pixel steps and were concerned with the effects on the SBF
pattern from any {\it ad hoc} interpolation scheme that might be required
to assemble the completed image from a random set of offsets.
Ironically, we did not use the WFC images in the present analysis,
given the excellent quality and superior resolution of the PC1 images;
however, this was a decision made after the data were in hand.

In the dither pattern,
each exposure would be shifted from the previous one, by $\sim10$ WFC pixels
($\sim1''$) in either the row or column direction, spiraling around
the original pointing, with the exact shift adjusted by $\pm1$ WFC
pixel to avoid having any object land on the same row or column
that it may have fallen on in a previous exposure.
This latter criterion also meant that the exposure was simultaneously stepped
$\pm1$ pixel in a direction perpendicular to the major step,
resulting in the spiral being skewed from perfect alignment with
WFC row and column axes.
The detailed shifts in all cases reflected the slight misalignment
of the WFC CCDs with respect to perfect quadrature with the {\it HST}
sky axes (see \markcite{holtzb}Holtzman et al. 1995b).

Integral pixel offsets for the WFC produce non-integral steps in PC1,
but for F814W, the PC1 images are nearly Nyquist-sampled,
so shifting the images to a common center may be done with little error.
As it is, however, we chose to stack the PC1 images with centering
only to the nearest pixel to avoid the complexity of patching
in the defects prior to interpolation.
While this produces a slight blurring
at the one-pixel scale, this has little effect on measurement of the SBF signal,
because the final composite image remains photon shot-noise limited
at the highest-spatial frequencies.
This can be simply understood by considering the enclosed energy curve
of the PSF.
\markcite{holtzb}Holtzman et al. (1995b) show that only 32\%\ of the light
within the F814W PSF falls within its central core, corresponding to only
a single photon for the typical exposure time in the present work.

\begin{figure}[tb]
\caption{The $710\times711$ pixel ($\approx32\arcsec$)
portion of the PC1 image centered on the BCG in Abell 262.  The
central portion of the montage shows the galaxy and its central dust
clouds; this region is masked from the SBF analysis.  The remainder
shows the residual after the galaxy model is subtracted (with
$8\times$ deeper stretch); most of the objects seen are globular
clusters.  Some dust is still visible and this is also masked for the
SBF analysis.}
\label{bcgpic1}
\end{figure}

\begin{figure}[tb]
\caption{The $771\times781$ pixel ($\approx34\arcsec$)
portion of the PC1 image centered on the BCG in Abell 3560.  The
central portion of the montage shows the galaxy and its central dust
ring; this region is masked from the SBF analysis.  The remainder
shows the residual after the galaxy model is subtracted (with
$40\times$ deeper stretch); most of the objects seen are globular
clusters.}
\label{bcgpic2}
\end{figure}

The final composite PC1 images were
assembled with an algorithm that looks at the statistical
properties of the data set at any pixel location, and rejects
extreme values as might be caused by cosmic ray hits or hot pixels.
The BCG images are shown in Figures \ref{bcgpic1} to \ref{bcgpic4},
with the galaxies themselves largely subtracted to emphasize fine detail.
Globular clusters are readily visible into the centers of all images.
This is the strength of using {\it HST} for SBF work --- globulars,
background galaxies, and dust clouds are easily recognized and excluded
when measuring the SBF power spectrum.
Dust clouds are also visible in three galaxies,
A262, A3560, and A3565;
indeed, their centers are completely or nearly obscured by dust.

\begin{figure}[tb]
\caption{The $791\times742$ pixel ($\approx33\arcsec$)
portion of the PC1 image centered on the BCG in Abell 3565.  The
central portion of the montage shows the galaxy and its central dust
lane; this region is masked from the SBF analysis.  The remainder
shows the residual after the galaxy model is subtracted (with
$60\times$ deeper stretch); most of the objects seen are globular
clusters.}
\label{bcgpic3}
\end{figure}

\begin{figure}[tb]
\caption{The $766\times766$ pixel ($\approx33\arcsec$)
portion of the PC1 image centered on the BCG in Abell 3742.  No dust
was seen in this image, so only the very central portions were
excluded from SBF analysis.  The concentration of globular clusters
around the galaxy is evident.}
\label{bcgpic4}
\end{figure}

Although SBF are measured from the PC1 images, we do need the WFC
images to measure the sky levels, which are required as part of the analysis.
Sky levels were measured from a $20''\times20''$ patch
extracted from the far corner of WF3.
This is the corner of WF3 diagonally opposite the WFPC2 vertex,
and is thus the portion of the WFPC2 field most distant from the BCG centers;
the typical displacement of the sky patch from the
galaxy centers is $\sim125''.$
The galaxies still contribute light to the sky measurements at this modest
distance; however, their contribution to the sky is readily estimated
and corrected for using ground-based surface photometry extending
to much larger radii.
The sky values given in Table \ref{tabsbf} have
been corrected for galaxy light contributions.

\begin{deluxetable}{rccccccccc}
\tablewidth{0pt}
\tablecaption{SBF Measures and Galaxy Distances}
\tablehead{
\colhead{Abell}&\f8bar&\colhead{$e^-$}&\colhead{Sky}&
\colhead{$(V{-}I)$}&\colhead{$(V{-}I)_0$}&\Mf8bar&\colhead{$(m{-}M)$}&\colhead{$
\pm$}&\colhead{GCLF }}
\startdata
 262& $33.35\pm0.15$&2.3&21.40& 1.401&1.317& $-0.642$ &33.77&0.19&25.85 \nl
3560& $32.33\pm0.10$&3.4&21.61& 1.289&1.234& $-1.183$ &33.35&0.14&25.12 \nl
3565& $32.47\pm0.08$&3.8&21.70& 1.286&1.239& $-1.151$ &33.47&0.13&25.72 \nl
3742& $33.08\pm0.08$&3.0&21.73& 1.305&1.270& $-0.950$ &33.88&0.12&25.77 \nl
\enddata
\label{tabsbf}
\tablecomments{SBF amplitudes are given as \f8bar\ magnitudes with
no reddening or k-corrections, or as number of $e^-$ detected.
Sky is $I_{814}$ magnitudes per square arcsecond, after correction
for galaxy-light contamination.  Colors are $(V-I)$ observed and
$(V-I)_0$ after reddening and k-corrections. \Mf8bar\ is the predicted
SBF absolute luminosity calculated from equation (2).
GCLF is the $I_{814}$ turnover magnitude of the globular cluster
luminosity function.}
\end{deluxetable}

\subsection{$(V-I)$ Colors}

As equation (\ref{eqM8bar}) shows, the \Mf8bar value for a given BCG
depends strongly on its $(V-I)$ color.
We measure $(V-I)$ from ground-based photometry over an
annulus between $5''$ and $15''$ in radius from the galaxy centers
to match the area of the PC1 field used for the SBF measurements.
The photometry for the three southern BCG was obtained
under excellent conditions at the CTIO 1.5m telescope.
For A262, the $I_{KC}$ image was obtained at the KPNO 4m, and
John Blakeslee kindly obtained the $V$ image at the MDM 2.4m.
For the small redshifts of the BCG, the K-corrections are
$K_V\approx2.0z,$ and $K_I\approx1.1z;$
galactic absorption corrections are $A_V:A_I:E(B-V)=3.04:1.88:1.00.$
The observed $(V-I),$ and reduced $(V-I)_0$ colors are both
given in Table \ref{tabsbf}.
In passing, we note that ground $(V-I_{KC})$ and WFPC2 $(V_{F555W}-I_{F814W})$
are essentially identical \markcite{holtza}(Holtzman et al. 1995a).
While we did not obtain WFPC2 F555W data, we did compare the $F814W$
photometry to the ground $I_{KC}$ data.
For $(V-I)\sim1.25,$ \markcite{holtza}Holtzman et al. (1995a) find
$I_{KC}-I_{F814W}\approx-0.04;$
unfortunately, even with this transformation, the WFPC2 $I_{F814W}$ fluxes
are still brighter than the ground values by 0.04 mag with a spread of 0.05 mag.
\markcite{a97}Ajhar et al. (1997) found excellent agreement between WFPC2
$I_{F814W}$ and ground $I_{KC}$ transformed to $I_{F814W}$ for their
sample, so the present mismatch is disappointing.
Since to first order, we might expect any deviations in ground $V$
to correlate with those in $I_{KC},$ we chose to base $(V-I)_0$
entirely on ground photometry rather than ground $V$ and WFPC2 F814W.

\subsection{Measurement of \f8bar\label{sbf}}

Measurement of the fluctuation signal was straightforward, following
the steps detailed in \markcite{t87}Tonry et al. (1997)
and \markcite{a87}Ajhar et al. (1997).
Once the PC images were cleaned of cosmic rays and stacked, we
subtracted sky; fitted elliptical profiles to the galaxy light;
generated and subtracted a synthetic galaxy;
ran DoPhot \markcite{sms93}(Schechter et al. 1993)
to detect stars, globular clusters, and background galaxies;
determined a luminosity function of these objects and created a mask
for removing objects brighter than a completeness cutoff; and then
performed the fluctuation analysis.  The basic step in measuring
fluctuations is to model the object's power spectrum as
\begin{equation}
\label{eqsbf}
P(k)=P_0\times E(k)+P_1,
\end{equation}
where $P_1$ is a constant giving the background level, and
$P_0$ is another constant used to scale the power spectrum of the PSF, $E(k).$
(Properly speaking, $E(k)$ is an expectation power
spectrum that incorporates the galaxy profile and the mask that has
been applied to the data.)

There are several sources of uncertainty in deriving the fluctuation power.
The first is the normalization and match of the PSF to the data.
The second arises in fitting the power spectrum: the lowest
wavenumbers are always corrupted, but there is no fluctuation signal
at high wavenumbers, so the choice of precisely which wavenumbers to
fit introduces uncertainty.  Finally, the raw $P_0$ variance is
corrected for the variance from residual, undetected point sources, and
this correction also carries an error.

We used the same PSF as Ajhar et al. (1997),
which was constructed from several F814W star images obtained in 1995.
The wings of the PSF were provided by archive exposures of stars centered in
PC1 to provide routine monitoring of the F814W filter zero point calibration.
The core of the PSF was constructed from four PC1 images of the
quadrupole gravitational lens 2237+0305, which were dithered in
a $2\times2$ pattern of 0.5 PC pixel steps.
This data set allowed the core to be recovered with
``perfect'' centering on a PC pixel.
The composite PSF has a total exposure of $5.8\times10^5$ photons.
It is an excellent match to the observed power spectra of the galaxies
with the highest signal to noise, A3560 and A3565, over the wavenumbers where
SBF dominates, so we deemed it to be suitable for all our observations.

We performed experiments with synthetic PSF images (created by software
developed by JAH) with calibrated amounts of defocus and miscentering.
We found that the
results are completely insensitive to where the PSF was centered, but
that $P_0$ became larger as the PSF became defocussed, as would be
expected from its more compact power spectrum.  Using a PSF from the
extreme secondary displacement caused by ``breathing'' of the OTA
quoted by STScI, we found that \mbar\ brightened by about 0.3 mag.
A synthetic PSF with perfect focus gave nearly the same results as
the observed PSF, but was still about 0.1 mag brighter, revealing a bit
{\it less} high wavenumber power than the observed PSF.  
Ajhar et al. also demonstrate that the observed PSF
will give very consistent results applied to a variety of
observations, and indeed, we are working differentially with respect to
those results anyway.
We thus conclude that the observed PSF is appropriate to use for the
fluctuation analysis, with mismatch and normalization error amounting
to about 0.05 mag.  

\begin{figure}[tb]
\plotone{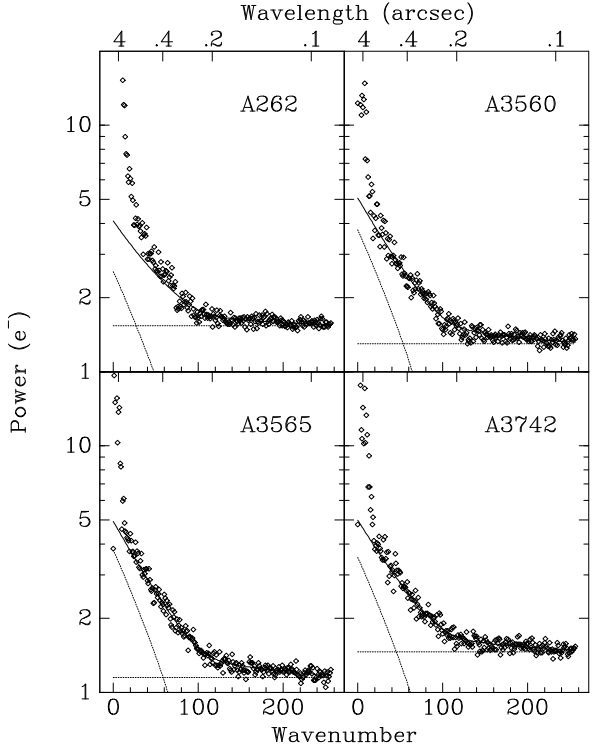}
\caption{The power spectra of each galaxy in the annulus with radii
5\s. 8--11\s. 7.  The dotted lines show the components of white noise
power and PSF power.  The solid lines show the sum of the fitted power
spectra.  The diamonds are the azimuthal averages of the data power
spectra.  The bottom axes indicate wavenumber; the top axes indicate
wavelength.}
\label{sbfpow}
\end{figure}

An observed power spectrum always has excess power at wavenumbers
below 10 or so (wavelengths longer than $\sim 30$ pixels),
arising from poor flattening, poor galaxy subtraction, dust, etc.
With the exception of A262 we found very good agreement with the PSF
beyond these contaminated wavenumbers.  We normally use
the rms variation in $P_0$ as a function of where we begin the power
spectrum fit as an additional error component.
In the case of A262 there is a range of $P_0$ values that are allowed by
the power spectrum; this range is reflected in the larger error
budget for this galaxy.
PSF fits to the power spectra of the BCG are shown in Figure \ref{sbfpow}.

The contribution from the residual point sources was small ---
always less than a 0.20 mag correction to $P_0$, and more
typically 0.10 mag.  Since we think we know this variance contribution
quite well, at least to 25\%, the contributed error is small.
As a test of this, we routinely analyze different annuli independently,
and except for A262 we used 4 annuli at 1, 2, 4, and 8\arcsec\ mean
radius (the central annulus was obscured by dust for A262).  If we
have an error in the residual point source correction (or an unknown
source of variance which does not scale with galaxy brightness) it
will show up as a radial gradient in \mbar.  A3560 had a gradient of
about 0.1 mag in each of the outer annuli, but the other three had no
gradient at the 0.05 mag level, giving us confidence that we have
modeled the residual variance well and we have no unforeseen source
of variance.  The gradient in A3560 may very well be real;
there is a 0.04 mag color gradient in $B-R$ from the center to $20''$
in radius, which would produce a gradient in \mbar\ consistent
with what we see.

The estimates and errors for \mbar\ are derived from the averages of
the values determined for each annulus, and the formal
uncertainty in the $P_0$ fitting procedure added in quadrature to 0.05
magnitude for the PSF uncertainty.  With the exception of A3560, where
we think we see a real gradient in \mbar, the scatter between the
different annuli is consistent with the error estimates.
We list the apparent SBF \f8bar fluxes in Table \ref{tabsbf}
uncorrected for extinction and prior to k-correction.
The SBF signal strength in electrons is also given.
A262 has the weakest SBF signal, which is reflected in its larger error bars.

\subsection{The Turnover of the Globular Cluster Luminosity Function}

\begin{figure}[tb]
\plotone{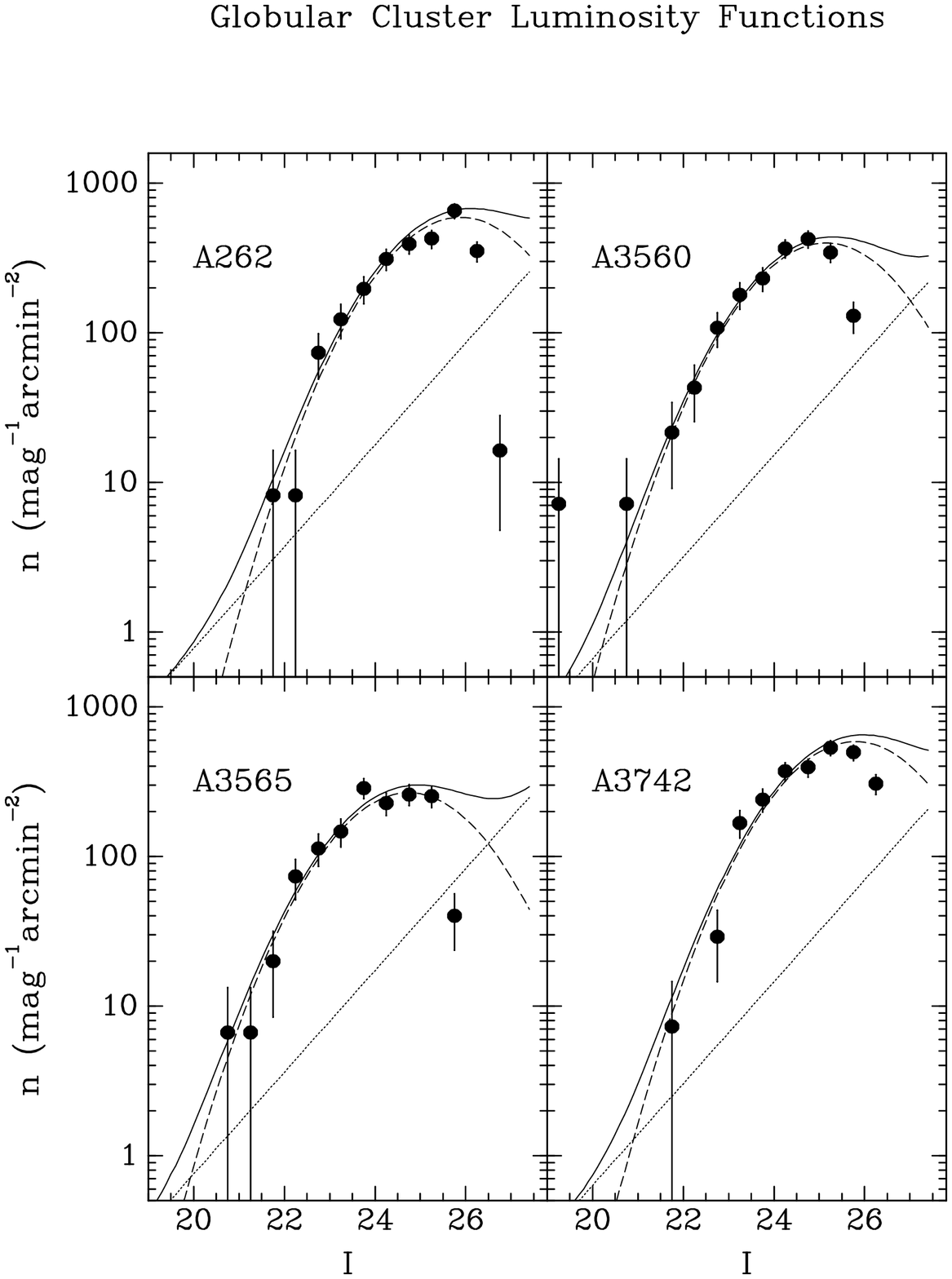}
\caption{The fitted GCLFs and background galaxy luminosity
functions.  The dashed lines are the GCLFs; dotted lines are the
background galaxies.  The solid lines are the combined luminosity
functions from which the residual variance is computed.  The error
bars only represent the Poisson scatter in each bin.  The radial
ranges shown are $3\s. 0 < r < 21\s. 4$, $1\s. 5 < r < 22\s. 5$,
$2\s. 1 < r < 23\s. 5$, and $0\s. 9 < r < 22\s. 1$
for A262, A3560, A3565, and A3742, respectively.}
\label{fgclfs}
\end{figure}

As discussed above,
measuring \mbar\ requires characterization of the galaxy's globular
cluster luminosity function (GCLF) to estimate the residual
variance contributed by the undetected (faint) portion of the GCLF as
well as by the undetected faint galaxies.
Figure~\ref{fgclfs} shows the luminosity functions of the
objects found in the images of our sample along with the GCLF,
background galaxy, and combined luminosity function fits.  The fitting
procedure naturally produces an estimate of the GCLF turnover
magnitude in the $I$ band for an assumed Gaussian $\sigma$ width of
1.4~mag.  The $m^0_I$ GCLF turnover magnitudes are listed in
Table~\ref{tabsbf}.  The estimated turnovers are all near the
$I \sim 25$ completeness limit and are uncertain by 0.2--0.3~mag.

Based on the SBF distance moduli, the mean absolute magnitude of the
four BCG GCLF turnovers in the $I$ band $\langle
M^0_I\rangle = -8.29\pm 0.18;$ the error is consistent
with the measurement errors
and an estimated intrinsic scatter in the GCLF technique of $\sim
0.25$~mag.  Few deep $I$-band GCLFs of giant elliptical galaxies have
been published for comparison; however, $\langle M^0_I\rangle$ is
consistent with the M87 measurements of \markcite{wslmb}Whitmore {\it
et al.} (1995), who found $m^0_{I, \rm M87} = 22.67$.  Using the SBF
distance modulus to the Virgo cluster of $31.03\pm 0.05$
(\markcite{t97}Tonry {\it et al.} 1997) yields $M^0_{I, \rm M87} =
-8.36$.  For additional comparison, a crude estimate of $M^0_I$ can be
based on the following assumptions: 1) Combining the distance modulus
to Virgo with the average observed apparent magnitude for the GCLF
turnover in Virgo in the $V$ band of $m^0_V = 23.75\pm 0.05$
(\markcite{bt96}Blakeslee \& Tonry 1996) yields $M^0_V = -7.28 \pm
0.07$.  2) Taking the mean color $\viz = 1.10\pm 0.1$, based on the
average of the mean \viz\ found in Coma's IC~4051 of $\viz = 1.08$
(\markcite{betal97}Baum {\it et al.} 1997) and that found in M87 of
$\viz = 1.12$ (\markcite{wslmb}Whitmore {\it et al.} 1995) and
applying an uncertainty of 0.1~mag, yields an estimate of $M^0_I =
-7.28 - 1.10 = -8.38 \pm 0.12$, also in good agreement with the BCG value.

\section{Measurement of $H_0$}

The present data set permits two separate approaches to measuring $H_0.$
The first and most obvious approach is simply to form Hubble ratios
for the four BCG and average them in an optimal way.
The BCG all have velocities in excess of 4,000 km s$^{-1}$ (in the CMB frame),
and may be far enough away so that their average Hubble ratio
might be close to the true value of $H_0.$
At the same time, this approach does not make use of the BCG Hubble
diagram, nor does it transfer the results to the 15,000 km s$^{-1}$ far-field.
The BCG instead are simply treated as test particles, without reference
to their photometric properties (although we do use cluster
averages for the velocities).

The second approach is to use the BCG as distance estimators to
avoid any use of the velocities of the four SBF BCG themselves,
in an effort to skip over the near velocity field --- this is
the approach promised by the title of our paper.
The frame independence of both the BCG Hubble diagram and $L_m-\alpha$
relationship argues that we can successfully reduce
the BCG to a common distance.
The SBF distances then permit calibration of the BCG
as absolute rather than relative distance estimators.

In either approach, the mathematical formalism used to derive a 
Hubble ratio is the same. For each SBF observation,
we compute a distance measurement in Mpc, $D$, from the
expression
\begin{equation}
\label{eqSBFdm}
D = dex(0.2 (\f8bar - \Mf8bar - 25))
\end{equation}
where \f8bar and \Mf8bar are corrected for
extinction and k-dimming.  The Hubble ratio is then just
\begin{equation}
\label{eqhratio}
H_0 = v / D
\end{equation}
where $v$ is either the observed velocity of the BCG in the appropriate reference
frame (for the first approach) or the estimated velocity of the
BCG using the prescription in Postman \& Lauer (1995) and summarized
in \S3.2 (for the second approach).

\subsection{$H_0$ from the SBF BCG Hubble Ratios Alone\label{ho4000}}

The BCG $(V-I)_0$ colors and the F814W SBF calibration of equation
\ref{eqM8bar} allow us to predict the absolute fluctuation magnitude \Mf8bar.
Using a \f8bar\ k-correction of $\approx 7z$
\markcite{t97}(Tonry et al. 1997) and our
measured values for the apparent \f8bar, we derive the distance
moduli and errors listed in Table \ref{tabsbf}.
Converting these to distances in Mpc (equation \ref{eqSBFdm}) 
and using the velocities in the CMB,
LG, and AC frames, we compute Hubble ratios in Table \ref{tabho}.
The errors listed here include the estimated velocity error of 184
km s$^{-1}$ plus a nominal 100 km s$^{-1}$ allowance for peculiar
velocity with respect to the local velocity field (added in quadrature),
although this term could plausibly be as large as 5\%.
There is no allowance for bulk flow since we are explicitly trying to remove
this by examining different reference frames.  The average Hubble
ratios are the weighted logarithmic averages, and the errors are those expected
given the individual distance errors.
 
While this sample is too small to solve for a preferred reference
frame at 4,000 km s$^{-1}$, we do see the insensitivity of the average
Hubble ratio among the three frames because of the sampling over the sky.
It is also apparent that $\chi^2$ prefers the CMB frame to the Local Group
or Abell Cluster frame, but again the numbers are small.
Because these SBF
distances appear to be extremely consistent and accurate,
a larger sample observed by {\it HST} throughout the sky could give
a precise measure of peculiar velocities;
we discuss this issue further in $\S\ref{flow}.$
Our preferred value for $H_0$ is derived from the CMB frame since it has
the lowest value of $\chi^2$: $H_0=82\pm8$ km s$^{-1}$ Mpc$^{-1};$
the error is discussed in detail below.
 
\subsection{The Far-Field $H_0$}

The premise behind seeking the far-field is that the near-field may be strongly
affected by random peculiar velocities or bulk flows.
One measures distances in the near-field, but avoids using
the corresponding near-field velocities by forming distance
ratios between near and far objects, and then forming a Hubble
ratio based purely on the far velocities.
The classic example of this approach is using the distance to the nearby Virgo
cluster, but adopting the velocity of the much more distant Coma
cluster by finding the ratio of the Virgo to Coma distance from
some form of relative distance estimator, such as the $D_n-\sigma$ method.

In the present case, the transference to the far field is somewhat less
obvious as we are not making an explicit comparison of the four SBF BCG to,
say, the most distant BCG in the \markcite{t97}Lauer \& Postman (1994)
sample at 15,000 km s$^{-1}.$
Instead, we will implicitly compare the SBF BCG to the entire
Lauer \& Postman sample on the presumption that all BCG can
be transferred to a common distance based on their observed velocities
and a simple linear Hubble flow model.
One can add a bulk flow to this model,
but in the end, this makes little difference, as we discuss below.

The BCG $L_m-\alpha$ distance estimator presented
in \markcite{lp92}Lauer \& Postman (1992, \markcite{t97}1994)
and \markcite{pl}Postman \& Lauer (1995),
indeed is based on comparing the physical properties of BCG
after adopting a linear Hubble flow and an {\it ad hoc} $H_0.$
The $L_m-\alpha$ relation works by allowing
$L_m$ to be predicted based on $\alpha$ as measured at the
metric radius $r_m.$
The scatter in $L_m$ is 0.24 mag, which translates into a typical
distance error of $\sim17\%;$
this is larger than the error expected for a pure inverse-square
distance estimator, as an error in the metric radius also
implies an error in the apparent luminosity within the metric radius
\markcite{go}(Gunn \& Oke 1975).
The $L_m-\alpha$ relation is equivalent to assuming that BCG
all have the same average enclosed surface brightness as a function
of physical radius for a given value of $\alpha.$
\markcite{pl}Postman \& Lauer (1995) showed that they could estimate
BCG redshifts (that is a relative distance expressed
as a velocity) to 17\% accuracy from a surface brightness curve
of growth by finding the angular aperture at which the enclosed
surface brightness and $\alpha$ were consistent with the
$L_m-\alpha$ relationship, then using the aperture size as a metric
distance estimator.

\begin{figure}[tb]
\plotone{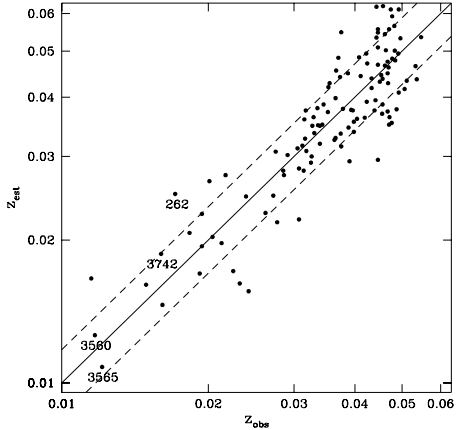}
\caption{Redshifts of the BCG estimated from the
$L_m-\alpha$ relationship versus
observed redshifts are shown (see Postman \& Lauer (1995)).
Lines indicate equality of the two redshift measures
and the $\pm1\sigma$ errors in the estimated redshifts.}
\label{zest}
\end{figure}

Figure \ref{zest} shows the plot of estimated versus observed redshift
presented by \markcite{pl}Postman \& Lauer (1995).
The estimated redshift results from calculating the velocity
required to bring a given BCG onto the ridge line of the $L_m-\alpha$ relation.
The scatter in $z_e$ thus reflects the scatter
of the BCG about the ridge line.
\markcite{lp}Lauer \& Postman (1994) showed that this scatter is strongly
dominated by random photometric differences between the BCG rather
than random peculiar velocities or velocity errors --- photometric
scatter is constant with redshift, as are the
residuals about the $L_m-\alpha$ relationship, while the decreasing
relative importance of random velocities with increasing Hubble velocity
would cause the scatter to decline with redshift.

The estimated redshifts for the four SBF BCG (given
in Table \ref{tabho}) imply the velocity
scaling required to bring their photometric properties onto the
$L_m-\alpha$ ridge line; 
Hubble ratios can then be calculated, based on the SBF distances.
The implied far-field Hubble ratios are given in Table \ref{tabho}.
Again, the average value is best estimated with a error-weighted logarithmic
average of the ratios.  The far-field Hubble constant implied
is $H_0=89\pm10~{\rm km}~{\rm s}^{-1}~{\rm Mpc}^{-1}.$
The error will be discussed in detail in the next section,
but in this case it is dominated by the BCG photometric
scatter about the $L_m-\alpha$ relationship.
In this context, it is worth noting that A262 is
among the most deviant BCG with respect to its position in the
$L_m-\alpha$ relationship, a conclusion echoed in the
highly deviant Hubble ratio that it yields using its photometric
redshift as the velocity (see the last column in Table \ref{tabho}).
The $\chi^2$ value for the far-field $H_0$ is also large.
Deleting A262, yields a lower far-field
$H_0=79\pm10~{\rm km}~{\rm s}^{-1}~{\rm Mpc}^{-1}.$
This value is consistent with the former value, but does
suggest that a more complete sampling of the nearby BCG
may be an attractive way to improve the accuracy of the BCG
far-field $H_0.$

As the far-field $H_0$ does not explicitly depend on the velocities
of the individual SBF BCG, it is explicitly independent of the velocity frame.
There may be {\it implicit} dependences on frame, since different
choices of Hubble velocity will affect the particular placement
of any given BCG within the $L_m-\alpha$ relationship;
however, such effects are tiny.
For example, at $\alpha=0.5,$ the ridge line value of $L_m$ varies
by only 1\% among the three frames.
More to the point, the $z_e$ varies by only 2\% for A262, and less
than 1\% for the other BCG over the various frames.
This result is not surprising, as any dipole pattern caused
by a large bulk flow superimposed on the BCG recession velocities
will integrate out of the $L_m-\alpha$ relationship because it
is defined from the full sky.

A more relevant question is whether the nearby SBF BCG set may be offset
from the more distant BCG by some global perturbation of the Hubble flow
that is limited to within, say, 5,000 km s$^{-1}$ or some other small
fraction of the 15,000 km s$^{-1}$ limit of the Abell Cluster sample.
This was the sort of problem posed by
\markcite{tco}Turner, Cen, \& Ostriker (1992)
and addressed by the \markcite{lp92}Lauer \& Postman (1992) BCG Hubble diagram.
The linearity of the BCG Hubble diagram argued that $|\delta_H|<0.07$
between 3,000 and 15,000 km s$^{-1}.$
Further, \markcite{lp92}Lauer \& Postman (1992) showed that beyond
9,000 km s$^{-1},$ which includes the bulk of the BCG sample,
$|\delta_H|\leq0.02$ on radial shells of 3,000 km s$^{-1}.$
In other words, there is no evidence that scatter in the
$L_m-\alpha$ relationship or its ridge line is affected
by any global variations in $H_0$ within 15,000 km s$^{-1}.$
The $L_m-\alpha$ relationship is defined by the full BCG sample,
unweighted by distance; the effective scale of the far-field
BCG Hubble constant is just the average BCG recession velocity,
or $\sim$11,000 km s$^{-1}.$

\subsection{Errors in $H_0$}

Both the ``SBF-alone'' and far-field $H_0$ values are affected by several
random errors.
The strong dependence of \Mf8bar on $(V-I)$ puts a premium on accurate colors.
We do not have multiple ground observations of the BCG,
but previous experience plus the quality of the standard star
solutions argues that the error in $(V-I)$ is likely to be close to 0.012 mag.
We also include a 10\%\ uncertainty in ${\rm E_{B-V}},$
and $\sim20$\%\ errors in the $(V-I)$ and \f8bar k-corrections.
The final net color error gives a typical $\sim0.09$ mag error
for estimating \Mf8bar from equation \ref{eqM8bar};
this is added in quadrature
to the measurement error in \f8bar given in Table \ref{tabsbf}
to give the total random distance error.
Calculation of the Hubble ratios also includes
a 184 km s$^{-1}$ velocity error and 100 km s$^{-1}$ peculiar velocity
term for the SBF-alone measures,
and a 17\% redshift error for the Hubble diagram values, as noted above.
The errors in the individual Hubble ratios given in Table \ref{tabho}
reflect the random error contributions only.

Important systematic errors include uncertainties in the F814W SBF calibration,
the absolute zero point of the $I_{KC}$ relationship on which it is based,
and errors in the {\it HST} PSF.
Scatter about the {\it HST} \Mf8bar versus $(V-I)$ relationship
argues that the uncertainty in predicting \Mf8bar, given perfect
$(V-I),$ is $\sim0.10$ mag \markcite{a97}(Ajhar et al. 1997).
\markcite{t97}Tonry et al. (1997) give $\sim0.07$ mag as the error
in the $I_{KC}$ SBF zero point, which itself is a composite of
the statistical errors in the ground $I_{KC}$ versus $(V-I)$ relationship
plus uncertainties in the Cepheid calibration of the relationship.
The systematic error associated with the PSF is 0.05 mag,
as discussed in $\S\ref{sbf}.$
If we add these errors in quadrature, we get a total systematic
uncertainty of 0.13 mag in predicting distances.

We include no systematic error in velocity for $H_0$ estimated
from the Hubble diagram.
For the SBF-alone $H_0$ estimate, however, we do include an error
for its relation to the true far-field value.
Since the four SBF BCG are among the nearest of the Lauer \& Postman (1994)
sample, we take the net uncertainty in $H_0$ measured at $\sim$4,500 km s$^{-1}$
scales as 7\%, given the variation of $\delta H\sim0.07,$
allowed out to 15,000 km s$^{-1}$
by the \markcite{lp}Lauer \& Postman (1992) BCG Hubble diagram.
In terms of velocity, this error corresponds to $\sim300$ km s$^{-1}$
at the SBF BCG distances, similar to plausible bulk flows
on this scale, regardless of how the BCG sample the
\markcite{lp}Lauer \& Postman (1994) flow.
The errors in the final average $H_0$ values given in Table \ref{tabho}
reflect the distance error, and except for the Hubble diagram solution,
the 7\% far-field error added in quadrature to the statistical error.

There are a number of paths for improving the present results.
The Hubble diagram $H_0$ can best be refined by obtaining
SBF distances to more nearby BCG.
Refining the $H_0$ estimated solely from SBF distances,
however, requires observation of more distant galaxies,
as the present use of near-field velocities is an important uncertainty.
Improving the F814W SBF calibration will also be useful.
Fortunately, {\it HST} SBF observations for the nearest galaxies
can generally be done within an orbit, and in many cases may be made
from images obtained for other purposes, as was the case for the
\markcite{a97}Ajhar et al. (1997) sample.
Because this calibration may improve,
we have attempted to make the path from \f8bar to distances readily visible.
Revised values of $H_0$ can thus be quickly derived when better data
become available.

\section{Discussion}

\subsection{Comparison of Recent $H_0$ Measurements}

The present $H_0$ values are somewhat larger than many
recent $H_0$ measurements based on calibration of
Hubble diagrams of a variety of distance estimators,
but they are consistent with the higher of the comparisons.
\markcite{gio97}Giovanelli et al. (1997) find
$H_0=69\pm5{\rm ~km~s^{-1}~Mpc^{-1}},$ using Cepheid distances to 12 galaxies,
which were used to calibrate a composite Tully-Fisher relationship
based on 24 clusters of galaxies within 9,000 km s$^{-1}.$
As do we, Giovanelli et al. emphasize their independence
from the classic Virgo/Coma distance ratio approach.
Giovanelli et al. did not show a Hubble diagram, nor list their
clusters, but from their description of the cluster distribution,
we estimate that the effective depth of their $H_0$
determination is $\sim$6,000 km s$^{-1}.$
Calibration of the Tully-Fisher relationship has been a
major goal of the {\it HST} Cepheid key project as well
\markcite{fre94}(Freedman et al. 1994).
\markcite{mou95}Mould et al. (1995) find $H_0=82\pm11{\rm ~km~s^{-1}~Mpc^{-1}},$
based on a Tully-Fisher calibration that reaches beyond
4,000 km s$^{-1},$ but that also depends heavily on the Virgo cluster sample.
\markcite{fre97}Freedman (1997) gives $H_0=73\pm8{\rm ~km~s^{-1}~Mpc^{-1}},$
as a provisional summary of the key project work to date.

Supernovae also are providing excellent probes of the Hubble flow.
\markcite{rpk}Riess, Press, \& Kirshner (1996) present a Hubble diagram based
on their light curve-shape method applied to 20 SN Ia's;
using three Cepheid calibrators, they conclude
$H_0=64\pm6{\rm ~km~s^{-1}}$~${\rm Mpc^{-1}}$ on $\sim$7,000 km s$^{-1}$ scales.
\markcite{ham}Hamuy et al. (1996) find $H_0=63\pm4{\rm ~km~s^{-1}~Mpc^{-1}},$
an essentially identical result, using their $\Delta m_{15}$
decay rate estimator applied to 29 SN Ia with an
effective depth of $\sim$12,000 km s$^{-1}.$
Proper use of SN Ia as distance estimators remains controversial, however,
with significant disagreement
on how the light curve decay rate relates to the SN Ia peak luminosity.
\markcite{sand96}Sandage et al. (1996), for example, emphasize a Hubble diagram
approach as well, using SN Ia's, and conclude
$H_0=57\pm4{\rm ~km~s^{-1}~Mpc^{-1}}.$
This value, while consistent with those of
\markcite{rpk}Riess, Press, \& Kirshner (1996) and
\markcite{ham}Hamuy et al. (1996),
argues for $H_0$ near the lower ends of their
error bars, rather than the upper ends, as would be more consistent with the
Tully-Fisher results cited above and the present BCG results.
\markcite{ken}Kennicutt, Freedman, \& Mould (1995) present a figure showing
large dispersion in $H_0$ values calculated from SN Ia alone over the
last few years; agreement on how to calibrate the SN Ia distance
scale remains work for the future.

There is an abundance of other recent measurements of $H_0$ that
we could cite, but as many of them are based heavily or exclusively
on the Virgo or Coma clusters, we find them less attractive
than methods featuring rich sampling of the Hubble flow.
It is also worth noting that while we are comparing $H_0$
estimates that all depend on the {\it HST} Cepheid calibration work,
different methods make use of different calibrators,
which may account for a portion of the variance in $H_0$ among authors.
At this writing, \markcite{fw97}Feast \& Whitelock (1997) are arguing
that the Cepheid scale, itself, should be revised based on {\it
Hipparcos} parallax measurements.
If so, then the \markcite{t97}Tonry et al. (1997) and
Ajhar et al. (1997) SBF zero points will need revision.

Two promising alternative approaches for measuring $H_0,$
which are completely independent of local calibration,
are gravitational lens induced time delay observations and measurement
of the Sunyaev-Zel'dovich (SZ) effect for $z >> 0.05$ clusters.
As \markcite{fre97}Freedman (1997) summarizes,
however, measurements of the SZ effect
are neither accurate nor consistent enough at this time to
provide an $H_0$ that challenges the more local measures.
Gravitational lenses also have the potential to produce
a far-field $H_0$ that steps over all the distance-ladder
problems that bedevil more traditional methods; however,
detailed mass distributions of the lenses are required,
which presently limits their accuracy.
\markcite{kun}Kundi\' c et al. (1997) find
$H_0=64\pm13{\rm ~km~s^{-1}~Mpc^{-1}},$
based on the observed time delay between the two QSO images
in the classic $0957+561$ lens at $z=0.36$.
In contrast, \markcite{s97}Schechter et al. (1997) present time delays and mass
models for the lens PG $1115+080$ that favor $H_0=42,$
although they also present a model that gives $H_0=64.$
It is thus difficult to make a strong case at present that
gravitational lenses, which probe the Hubble flow on extremely
large scales, are yielding consistent $H_0$ values significantly
smaller than the $z\leq0.05$ measurements.

\subsection{Is the BCG Far-Field Far Enough?\label{far_enough}}

Observational evidence suggests that the far-field has been reached.
While the sampling of the Hubble flow remains sparse beyond 15,000 km s$^{-1},$ 
the limiting radius of \markcite{lp92}Lauer \& Postman (1992),
the SN Ia Hubble diagrams of \markcite{rpk}Riess, Press, \& Kirshner (1996) and
\markcite{ham}Hamuy et al. (1996) are consistent with linear flows out
to $\sim$30,000 km s$^{-1}.$
Going to cosmological distances,
Kim et al. (1997) use SN Ia to constrain $\delta_H$ by comparing
28 SN with $0.35<z<0.65$ to 18 SN of the Hamuy et al. (1996) sample.
The motivation for developing such SN Hubble diagrams is to
measure $\Omega_M$ and $\Lambda;$ such cosmological tests presume
an unbiased local Hubble flow as a point of departure.
On the other hand, with assumed $\Omega_M$ and $\Lambda,$
one can test for significant $\delta_H$ over extremely large scales.
For $\Omega_M\leq1,$ Kim et al. find $\delta_H<0.05$ ($1\sigma),$
or $\delta_H<0.1$ (95\%\ confidence).
For $\Omega_M<<1$ with $\Lambda=0,$ or 
$\Omega+\Lambda=1,$ one can get $\delta_H\lesssim-0.1,$
corresponding to global $H_0$ actually {\it larger} than the local value.

Going out far enough to measure the unbiased $H_0$ means reaching the
scale on which mass density fluctuations no longer generate
significant velocity perturbations of the Hubble flow.
The most important bias to consider for a full-sky
determination of $H_0$ is the global
radial retardation or acceleration of the
Hubble flow that occurs within significant mass over- or under-densities.
The possibility that we are within a large bubble of lower than
cosmic density, for example has been proposed as a way of reconciling
apparently high $H_0$ values measured nearby, with the estimated
age of the universe, and the often-cited concern (e.g. \markcite{bar}Bartlett
et al. 1995) that extremely far-field $H_0$ measures such as those from
the SZ effect or lenses are lower than the more local measures
(although, as noted above, the case for this is weak).
\markcite{ber}Berschinger (1985) and \markcite{ryd}Ryden (1994) present analytic
treatments of how voids grow with time.
The voids effectively expand faster than the cosmic scale factor;
observers well inside the voids would see linear,
if spuriously rapid Hubble flows.
\markcite{shi}Shi, Widrow, \& Dursi (1996)
in general find $\delta_H\sim-0.6\delta M/M,$
where $\delta M/M$ is the relative mass-deficit of the void.

When limited to popular initial power-spectra, however, one predicts
only small $\delta_H$ over the large volumes sampled by BCG and SN Ia.
\markcite{tco}Turner, Cen, \& Ostriker (1992), for example,
confined their analysis
to very modest scales compared to the Hubble diagrams now available.
The large variations in $\delta_H$ that they observed under
CDM and PIB power-spectra occurred for volumes limited to 3,000 km s$^{-1}.$
For volumes limited to 6,000 km s$^{-1},$ they found
$\langle\delta_H\rangle<0.05,$ depending on the power-spectrum.
\markcite{shi}Shi, Widrow, \& Dursi (1996) likewise conclude that
$\langle\delta_H\rangle\sim0.05$ on 15,000 km s$^{-1}$
scales with ``reasonable'' models of galaxy and structure formation.

The question of whether or not we have gone out far enough
in measuring the Hubble flow thus remains an issue only
if there is significantly more power in mass-fluctuations
on large scales than would be expected under standard theories.
In this context, we can posit that the 689 km s$^{-1}$ bulk flow
observed by \markcite{lp}Lauer \& Postman (1994) in the volume limited
to 15,000 km s$^{-1}$ is indeed evidence for such power on large scales.
\markcite{s95}Strauss et al. (1995),
and \markcite{fw}Feldman \& Watkins (1994) both find
the Abell cluster bulk flow to be incompatible at $\sim95\%$ confidence
with all standard models of galaxy formation considered.
\markcite{teg}Tegmark, Bunn, \& Hu (1994) further argue that the Lauer-Postman
bulk flow is not compatible with degree-scale measurements
of the CMB anisotropy power-spectrum.
Even so, however, \markcite{shi}Shi, Widrow, \& Dursi (1996), taking the
Harrison-Zel'dovich power spectrum shape and normalization
parameters that best fit the Lauer-Postman flow
(see \markcite{jk}Jaffe \& Kaiser 1995),
still find that only modest $\delta_H=0.05$ would be expected.
At the same time, Shi, Widrow, \& Dursi show that the range
of power spectra considered by Jaffe \& Kaiser to fit the Lauer-Postman
bulk flow would admit $\delta_H$ as large as 0.12 in the limiting
extreme case.

Of course, power-spectra give only a statistical expectation
for the local distribution of matter.  One remains free to
argue that we are within a local density anomaly
that exceeds the $\pm1\sigma$ fluctuations at some level.
In this case, \markcite{shi}Shi, Widrow, \& Dursi (1996)
discuss the CMB dipole and quadrupole anisotropies
that would be observed within the anomaly and conclude
that they would put strong constraints on the allowed geometry
of the local density fluctuation (see also \markcite{tom}Tomita 1996).
We conclude that while one can construct models of the universe
for which the far-field remains at distances well in excess of those
explored here, they are extremely unfavored by what we know of
the power-spectrum of initial mass fluctuations in the universe.

\subsection{The Abell Cluster Bulk Flow\label{flow}}

The accuracy of these SBF distances offers the means to test the
validity of different reference frames, although we cannot
legitimately solve for an independent, best-fit reference frame with
only four points.  As described in section $\S$\ref{ho4000}, we
consider three reference frames: CMB (for obvious reasons), Local
Group (since for smooth flows this has zero dipole locally), and Abell
Cluster (AC) frame following the \markcite{lp}Lauer \& Postman (1994)
bulk flow.  Although we tried to remove all systematic, common
contributors to the errors in the SBF distances when calculating
$\chi^2$ for our $H_0$ estimates, we find that $\chi^2/N$ in Table
\ref{tabho} ($N=3$) has a value of 0.3 for the CMB frame.  This may
indicate that we have overestimated our errors, since such a value or
lower will occur only 18\% of the time by chance, so we bear in mind
that all the true $\chi^2$ values may be larger than what is listed in
Table \ref{tabho}.  At the same time, it seems unlikely that our
random errors could be overestimated by a factor of almost 2,
necessary to raise $\chi^2/N$ in the CMB frame to unity, so at least
some of the concordance of results in this frame surely is
coincidence.  Taking the numbers at face value, the probability that
$\chi^2/N$ is at least as large as 1.2 (the LG value) is 0.31, and the
probability that $\chi^2/N$ is at least as large as 2.4 (the AC frame
value) is 0.07.  Thus our observations offer the most support for
these four clusters being at rest in the CMB frame.
 
\begin{deluxetable}{rrrrrrr}
\tablecolumns{5}
\tablewidth{0pt}
\tablecaption{Hubble Ratios and $H_0$}
\tablehead{
  & \colhead{D} &  &
\colhead{$H_0$} & \colhead{$H_0$} & \colhead{$H_0$} & \colhead{$H_0$} \\
\colhead{Abell} & \colhead{Mpc} &\colhead{$z_{est}$} &
\colhead{(CMB)} & \colhead{(LG)} & \colhead{(AC)} & \colhead{(BCG)}}
\startdata
 262 &$57\pm5$&0.0250&$82\pm8$&$90\pm9$&$94\pm9$&$132\pm25$\nl
3560 &$47\pm3$&0.0126&$86\pm7$&$75\pm7$&$72\pm6$&$ 81\pm15$\nl
3565 &$49\pm3$&0.0108&$83\pm6$&$73\pm6$&$70\pm6$&$ 66\pm12$\nl
3742 &$60\pm3$&0.0187&$78\pm6$&$80\pm6$&$79\pm6$&$ 94\pm17$\nl
\tablevspace{5pt}
Averages & & &     $82\pm8$&$79\pm8$&$78\pm8$&$89\pm10$\nl
\tablevspace{5pt}
$\chi^2/N$ & &       & 0.3& 1.2&  2.4&  2.6   \nl
\enddata
\label{tabho}
\tablecomments{The third column is the
redshift estimated from the BCG $L_m-\alpha$ relationship
(see Postman \& Lauer 1995).
Hubble ratios are in km s$^{-1}$ Mpc$^{-1}.$
The ``Averages'' line gives the logarithmic average of the individual ratios,
with error bars reflecting all systematic errors (see text).
$\chi^2/N$ is calculated using errors without common, systematic contributions,
and gives an indication of the internal consistency of the ratios.}
\end{deluxetable}

The very small number of SBF distances does not permit us say whether
this rejects the AC frame or not, since the AC frame was chosen to
minimize the scatter of the $L_m-\alpha$ relationship for 119 clusters
at distances much greater than our sample here.  Additionally, we are
guessing at a random velocity component of 209 km s$^{-1}$ (100 and 184 in
quadrature).  If the random velocity is as large as 500 km s$^{-1}$ the AC
frame will have unity $\chi^2/N$ (0.1 in the CMB frame).
 
It is clear that A262 is problematic for $L_m-\alpha$ as seen in
Figures \ref{bcg_hub}, \ref{bcg_hub_frame}, and \ref{zest};
it has a very low surface brightness and hence
its distance is estimated to be large.
Likewise, the SBF and
$L_m-\alpha$ distances are consistent within the errors for A3560,
A3565, and A3742, but there is a $3\sigma$ inconsistency with A262.
The cosmic scatter in the $L_m-\alpha$ is clearly related
to intrinsic variations in metric surface brightness among the BCG.
Lauer \& Postman (1994) note that this scatter greatly dominates
any variation in $L_m$ due to peculiar velocity,
but rely the assumption that intrinsic variations
in BCG surface brightness are not correlated over large angles
and will average out of a large sample.
In the case of A262, however, the AC frame does partially compensate
for the discrepancy between its $L_m-\alpha$ distance and its redshift.
The rms scatter in the $L_m-\alpha$ relationship for just
these four clusters is 0.41 mag in the CMB frame, but only 0.27 mag in
the Abell Cluster frame.  If we delete A262 the rms $L_m-\alpha$
residuals drop to 0.29 mag (CMB), and 0.20 mag (AC).  In their
analysis \markcite{lp}Lauer \& Postman tried deleting outliers (such
as A262) and found the AC frame to be quite stable, so this is not the
whole story.  Nevertheless it seems clear that a larger sample of SBF
distances could not only provide a very accurate reference frame and
random velocity amplitude, but also allow us to understand why the
Lauer \& Postman sample and the $L_m-\alpha$ relation point to the AC frame.

\subsection{The Distance to the Virgo Cluster}

\markcite{lp92}Lauer \& Postman (1992) originally attempted to calibrate
the BCG Hubble diagram on the presumption that NGC 4472,
the Virgo cluster BCG, was typical.
Adopting a 14.4 Mpc distance to NGC 4472, they found
$H_0=77\pm8{\rm ~km~s^{-1}~Mpc^{-1}};$ however, if they adopted the
\markcite{sand90}Sandage \& Tammann (1990) 21.9 Mpc Virgo distance for NGC 4472,
$H_0=51\pm5$ would be implied.
Our present results for $H_0$ are essentially identical to
that obtained with the short distance to NGC 4472, and are significantly
different from that obtained from the with distance, arguing
that the distance to the Virgo core is indeed close to the shorter value.
If NGC 4472 were at the distance of 21.9 Mpc, then it would
be among the brightest of the \markcite{lp}Lauer \& Postman (1994) sample,
deviating from the $L_m-\alpha$ ridge line by more than $2\sigma.$
Of course, one could question this conclusion by challenging
the SBF calibration, since it is already known that SBF distances
imply a short distance to Virgo \markcite{t97}(Tonry et al. 1997);
however, this still remains as an important consistency check.
The relative distances inferred from BCG are consistent with SBF distances.

\section{Conclusion}

We have used {\it HST} to obtain SBF distances to four BCG beyond
4,000 km s$^{-1}$ to calibrate the \markcite{lp92}Lauer \& Postman (1992) BCG
Hubble diagram, producing an estimate of the global value
of $H_0$ valid on $\sim$11,000 km s$^{-1}$ scales.
This method gives $H_0=89\pm10{\rm ~km~s^{-1}~Mpc^{-1}},$
and is based on the full \markcite{lp}Lauer \& Postman (1994) 15,000 km s$^{-1}$
volume limited BCG sample.
As such, the result is independent of Virgo or Coma cluster distances and
membership issues, as well as the recession velocities of the four BCG studied.
The large error reflects the photometric scatter about the
$L_m-\alpha$ ridge line, which was used to transfer the BCG Hubble
diagram to the SBF distance scale.
As more BCG are observed with {\it HST,} the formal errors
in this far-field $H_0$ should decrease.

Our review of the present understanding of the formation of
large scale structure argues that we are likely to
have fairly sampled the far-field.
Even theories with enough power on large spatial scales to generate bulk
flows as large as those observed by \markcite{lp}Lauer \& Postman (1994)
are unlikely to have deviations outside of $|\delta_H|\lesssim0.05$
for the volume sampled by the BCG Hubble diagram.
In contrast, the compatibility of our results with those based on more nearby
objects argues that there is little effect on $H_0$ and the depth
of the measurements.  Going to the far-field most likely
removes a source of uncertainty, rather than correcting for a systematic error.
Indeed we find $H_0=82\pm8{\rm ~km~s^{-1}~Mpc^{-1}}$
just from Hubble ratios based on the SBF distances and observed
recession velocities to the
four SBF-calibrated BCG at $\sim$4,000 km s$^{-1}$ alone,
a result consistent with our far-field result.

The present $H_0$ rests on calibration of the SBF method and
an understanding of its systematic effects.
At the fundamental level, we are tied to the nearby Cepheid calibrators.
Changes in the Cepheid scale will propagate to the present
results through the Tonry et al. and Ajhar et al. calibrations.
As noted in the introduction, \markcite{t97}Tonry et al. (1997)
SBF calibration is tied to seven spiral galaxies with Cepheid distances.
Further, Tonry et al. have observed enough galaxies to perform
an exhaustive series of tests, finding no systematic offsets
between SBF observations of bulges and elliptical galaxies.
A weaker link is transferring the ground $I_{KC}$ method to
the WFPC2 F814W filter, a task accomplished by
\markcite{a97}Ajhar et al. (1997);
we will attempt to refine this calibration as more nearby systems
are observed with {\it HST.}
We conclude that the major uncertainties in the distance
scale are those close to home rather than far away.

\acknowledgments

We thank Guy Worthey and Barbara Ryden for helpful
discussions, and John Blakeslee for the photometry of A262.
This research was supported in part by
{\it HST} GO analysis funds provided through STScI grant GO-05910.03-94A.

\end{document}